# Introducing Network Coding to RPL: The Chained Secure Mode (CSM)

Ahmed Raoof[*]
Dep. of Systems and Computer Eng.
Carleton University, Canada
Email: ahmed.raoof@carleton.ca

Chung-Horng Lung[*]
Dep. of Systems and Computer Eng.
Carleton University, Canada
Email: chlung@sce.carleton.ca

Ashraf Matrawy[*]
School of Information Technology
Carleton University, Canada
Email: ashraf.matrawy@carleton.ca

*Abstract*—The current standard of Routing Protocol for Low Power and Lossy Networks (RPL) incorporates three modes of security: the Unsecured Mode (UM), Preinstalled Secure Mode (PSM), and the Authenticated Secure Mode (ASM). While the PSM and ASM are intended to protect against external routing attacks and some replay attacks (through an optional replay protection mechanism), recent research showed that RPL in PSM is still vulnerable to many routing attacks, both internal and external. In this paper, we propose a novel secure mode for RPL, the Chained Secure Mode (CSM), based on the concept of intra-flow Network Coding. The main goal of CSM is to enhance RPL's resilience against replay attacks, with the ability to mitigate some of them. The security and performance of a proof-of-concept prototype of CSM were evaluated and compared against RPL in UM and PSM (with and without the optional replay protection) in the presence of Neighbor attack as an example. It showed that CSM has better performance and more enhanced security compared to both the UM and PSM with the replay protection. On the other hand, it showed a need for a proper recovery mechanism for the case of losing a control message.

## I. INTRODUCTION

Made into a standard in 2012, RPL [1] has attracted a great deal of research interest. In particular, routing security in RPL was of special interest, including different routing attacks the protocol is susceptible to [2], [3], mitigation methods and Intrusion Detection Systems (IDSs) [4], [5], and performance evaluation of some of RPL's security mechanisms [6]–[8].

Raoof *et al.* in [6], [7] showed that RPL's secure modes, while providing reasonable mitigation of some external attacks, are still vulnerable to many routing attacks (both internal and external) - see §II-A. In this paper, we propose a novel secure mode for RPL - the Chained Secure Mode (CSM) - which is designed using the principle of intra-flow Network Coding (NC) [9], [10] to introduce an extra layer of security for RPL's communications and to provide RPL with mitigation capabilities against several routing attacks, while keeping the same working principles of RPL.

Our contributions can be summarized as follows:
- We designed a novel secure mode for RPL, the CSM. This new secure mode uses the principle of intra-flow NC to create a linked chain of coded RPL control messages between every two neighboring nodes. The chaining effect can limit adversaries' ability to launch routing attacks, e.g., Wormhole, identity-cloning, or RPL-specific attacks such as replay or Neighbor attacks [2].
- A proof-of-concept prototype of the proposed CSM was implemented in Contiki Operating System (OS) [11].
- A security and performance comparison between RPL in CSM (as a prototype) and PSM against the Neighbor attack (NA) was conducted using several metrics. The results showed that CSM is capable of mitigating the attack with less overhead and power consumption than PSM with replay protection.

## II. THE PROPOSED CHAINED SECURE MODE (CSM)

### A. Motivations

RPL standard offers three modes of security [1], [12] to ensure control messages' confidentiality and integrity: ***UM***, where only the link-layer security is applied, if available (default mode); ***PSM***, which uses preinstalled symmetrical encryption keys to secure RPL control messages; and ***ASM*** uses the preinstalled keys to let the nodes join the network, after that all routing-capable nodes must acquire new keys from an authentication authority. In addition, RPL provides an optional replay protection mechanism that employs the use of Consistency Check (CC) messages [1], only available in the preinstalled (PSMrp) or authenticated mode (ASMrp).

The authors in [6], [7] have shown that PSM is able to mitigate most of the external attacks[1], while it does not enhance RPL's security against the internal attacks[2]. Furthermore, their work showed that external adversaries still can launch replay attacks, even when PSMrp is used (e.g., in the case of the Wormhole attack.)

A further investigation of RPL standard [1] shows that it only provides confidentiality and integrity of its control messages, without any verification of their authenticity. This opens the door wide open for attacks such as message-forging, identity-cloning, eavesdropping, and replay attacks [2] to be launched regardless which RPL secure mode is running. For

---
[1]An external attack refers to an attack that is launched by an adversary who is not part of the network, e.g., it does not have the encryption keys used by the legitimate nodes for RPL in PSM, or runs RPL in UM.

[2]An internal attack is launched by an adversary who is part of the network, e.g., it has the encryption keys used by the legitimate nodes for RPL in PSM.

[*] The authors acknowledge support from the Natural Sciences and Engineering Research Council of Canada (NSERC) through the Discovery Grant program.

978-1-7281-8326-8/20/$31.00 ©2020 IEEE

example, an external adversary can launch a ***Neighbor attack***[3] by merely monitoring the "*Type*" and "*Code*" header fields in any Internet Control Message Protocol (ICMPv6) message to identify RPL's DIO messages[4], without the need to decrypt the actual message [6].

The lack of message authenticity in RPL motivated us to overcome this problem, and NC came into the light as a possible solution. Incorporating the intra-flow NC into RPL provides any receiving node with a proof of message authenticity, assuming that the first message came from the original sender. This case stands true for most attacks as the adversaries normally join the network after it has been initiated and stabilized.

*B. How CSM Operates*

The idea behind CSM is to integrate the message chaining effect of intra-flow NC into RPL by *encoding / decoding* the current control message with a Secret Chaining (SC) value that is sent within the previous control message. These SC values are 4 bytes unsigned integers, locally unique per neighbor, and randomly generated for each control message.

Since RPL sends its control messages as either an Multicast (MC) or Unicast (UC) messages, CSM considers them as two independent flows: an MC-flow and a UC-flow. Because of that, every node in the network should maintain a table (the *SC table*) of the following SC values for each neighbor, in order to successfully encode and decode their control messages:

1) **SC_UC_RX:** The SC value used to decode the next incoming UC-flow message from the neighbor.
2) **SC_MC_RX:** The SC value used to decode the next incoming MC-flow message from the neighbor.
3) **SC_UC_TX:** The SC value used to encode the next outgoing UC-flow message to the neighbor.

In addition, each node should maintain the next SC value for its next MC-flow transmission (**SC_MC_TX**). For simplicity, the current proof-of-concept design uses *zero* as a value for the SC used for the first transmission in each flow.

To exchange the SC values used to encode the next control message, CSM employs the *RPL Control Message Options* from the standard [1]. These optional add-ons are used to provide (or request) information to the receiver(s). CSM adds two new options to accommodate the transmission of the next SC used for each flow: the (**SC_UC_NEXT**) option includes the SC value to be used for the next UC-flow message, and (**SC_MC_NEXT**) is for the SC value to be used for the next MC-flow message.

When a node wants to send an RPL control message (whether for the UC- or MC-flow), it will prepare the message as per the standard PSM procedures. However, two additional steps are performed by CSM before encrypting the message with the preinstalled key:

---

[3](an attack where the adversary replays any DODAG Information Object (DIO) messages it hears without modification, deceiving the victim nodes into thinking that the original sender is within their range)

[4](Code = 1 or 129) ⇒ a regular or secure DIO message, respectively.

- The *Code* field of the ICMPv6 header is encoded using the corresponding SC_UC_TX or SC_MC_TX value to mitigate the security vulnerability addressed in §II-A.
- Adding the (SC_UC_NEXT) and (SC_MC_NEXT) new control message options, as per the RPL standard. CSM should add both options for UC-flow messages and only the (SC_MC_NEXT) for the MC-flow messages. The use of both options for the UC-flow allows for quicker recovery from message chain breakage in the MC-flow.

After encrypting the message (according to standard PSM procedures), CSM will encode the whole message using the corresponding SC value then send it as usual. Fig.1 depicts how CSM constructs an RPL message.

At the receiving node, the decoding SC value is found from the SC table using the sender IP address. Then, it is used to decode both the *Code* field of the ICMPv6 header (to identify the type of RPL message) and the whole message, which is then processed as per PSM procedures. Any messages with a non-decodable *Code* field are discarded without processing.

Except for the above-mentioned modifications, CSM follows the same rules as in the RPL PSM standard.

## III. EVALUATION OF THE CHAINED SECURE MODE

To evaluate our proposed CSM, we conducted a security and performance comparison between our devised prototype of CSM and the currently implemented secure modes: RPL in UM, PSM, and PSMrp (both according to [12]). All the secure modes were evaluated in both normal operation and with an external adversary launching a Neighbor attack [2] (as an example of replay attacks – see §II-A).

Cooja, the simulator for Contiki OS [11], was used for all the simulations (with simulated motes). Fig.2 shows the topology used in our evaluation. A list of simulation parameters is provided in Table I. The simulations' results were averaged over ten rounds per experiment with a 95% confidence level.

Our evaluation uses the following metrics of the average: data packet delivery rate (PDR), data End-to-End (E2E) latency, number of exchanged RPL control messages, and network power consumption per received data packet.

The following assumptions were used: only the legitimate nodes send data packets toward the root (1 packet/minute per node). RPL is set up with the default Objective Function (OF) – the Minimum Rank with Hysteresis Objective Function (MRHOF) [13], while Contiki OS is using its default uIP stack (similar to [7]) . Also, we assumed no Link-layer security measures or encryption are enabled.

For the adversary, it operates in the same RPL secure mode as the legitimate nodes, but without the required preinstalled encryption key (for PSM, PSMrp, and CSM experiments). The adversary starts as a legitimate node, tries to join the network, then launches the attack after two minutes.

## IV. RESULTS AND ANALYSIS

*A. Analysis of the Results*

**Effects on the data packet delivery rate (PDR):** Looking at Fig.3a, it is clear that PSMrp and CSM successfully elim-inated the Neighbor attack effect, with both of them having almost 100% PDR. UM suffered the most (PDR ≈ 80%) as the adversary was able to become part of the network, while PSM was affected by a small margin (PDR ≈ 92%) as the adversary affected only one node (node 5) when it replayed the DIO messages it heard from nodes (1 and 2).

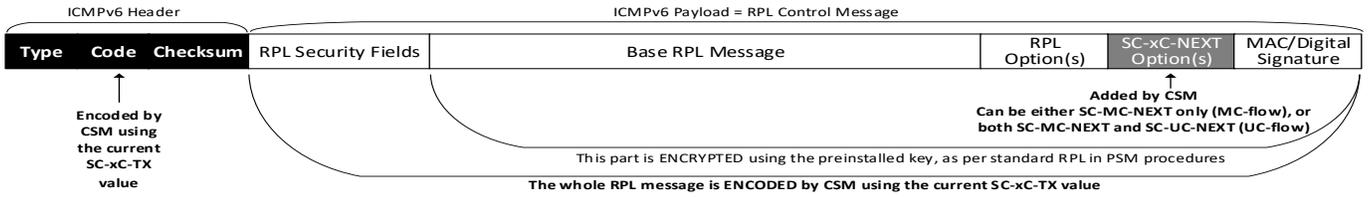

Fig. 1. Format of an RPL control message, as constructed by the proposed CSM. The black parts represents ICMPv6 header, the white parts are standard RPL in PSM fields, and the grey part is added by CSM. **Bold** text also presents additional steps performed by CSM.

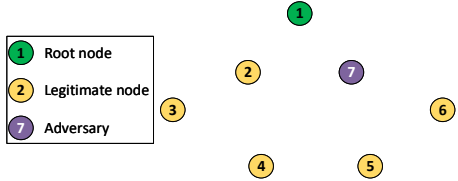

Fig. 2. Network topology used for the evaluation.

TABLE I
LIST OF SIMULATION PARAMETERS

| Description | Value |
|---|---|
| No. of scenarios | Two (No attack + Neighbor attack) |
| No. of experiments per scenario | Four (UM, PSM, PSMrp, and CSM) |
| No. of sim. rounds per exp. / time | 10 rounds / 20 min. per round |
| Deployment area | 60m W x 85m L |
| Number of nodes | 7 (adversary included) |
| Sensor nodes type | Arago Sys. Wismote mote |

**Effects on the data E2E latency:** as pointed out in [6], [7], Fig.3b confirms that the Neighbor attack introduces higher E2E latency to the network. This is clear in the cases of UM (latency ≈ 25 sec.) and PSM (latency ≈ 5 sec.). On the other hand, both PSMrp and CSM were able to mitigate the attack and kept the latency to its minimum (in the milliseconds).

**Effects on the exchanged number of control messages:** As seen in Fig.3c, the number of control messages sent in the network is almost the same for all the secure modes, with the attack increasing the number slightly. Under the attack, nodes running PSM are receiving more control messages than the other secure modes, due to the many MC DIO messages from the nodes 5 and 6 to the "ghost" parents (nodes 1 and 2). PSMrp nodes had a bit more control messages received when the Neighbor attack is commenced, compared to the no-attack scenario, due to the exchange of the CC messages.

On the other hand, our CSM prototype has the least number of received control messages, even less than what it had been sent originally. It was observed that this is due to some unicast Destination Advertisement Object (DAO)/DAO Acknowledgement (DAO-ACK) messages being lost (e.g., lossy wireless connections), which broke the UC message flow and resulted in having less received control messages than the sent.

It is worth noting that the number of received control messages is always higher than the sent one because many of the sent control messages are multicast messages which will be received by all neighboring nodes of the sender.

**Effects on power consumption:** We can see from Fig.3d that the power consumption patterns for RPL in UM, PSM, and PSMrp are very similar, with the attack slightly increasing the power consumption due to the undelivered data packets. However, it is noticeable that our CSM prototype is using less power than the other modes. From our observation, this behavior is because of the dropped control messages (whether the replayed messages or due to the message chain breakage).

### B. Observations

Some observations from the experiments include:

*1) Enhanced Security Features of CSM:* Those can be summarized as follows:

i) CSM adds an extra layer of security by encoding the control messages and chaining them with the SC values, which limits the adversaries' ability to eavesdrop on, manipulate, forge, and replay RPL control messages.
ii) Encoding of the *Code* field at the ICMPv6 header in CSM prevents external adversaries from identifying the type of RPL control messages, except for the first message of each flow as it is encoded with zero. Hence, message-type-specific external replay attacks (e.g., the Neighbor attack) can be mitigated by using CSM.
iii) The PSMrp mitigates only "one-way" replay attacks, which replay control messages from a node but not their correspondence. This proved to be inefficient with "two-way" replay attacks such as Wormhole attacks [6]. Since CSM uses control messages chaining (by the SC values) as a message authentication mechanism, any message encoded with unknown SC value will be dropped without an extra a challenge/response mechanism as in PSMrp.

*2) CSM Reduction of the In-threat Period:* The in-threat period can be defined as "*the time period in which an adversary can overhear, fully or partially understand the exchanged RPL control messages, and launch attacks*". This period ranges between **zero** (*the adversary cannot launch*

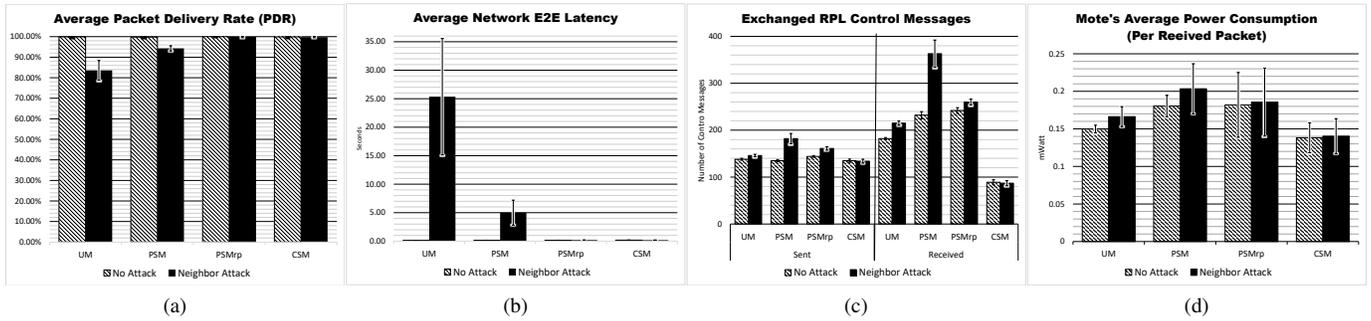

Fig. 3. Simulation results for the two scenarios and the four experiments.

*attacks successfully*) to **infinity** (*the adversary can launch attacks at any time*), depending on the secure mode used, the adversary type, and the attack.

For UM, the in-threat period is **infinity** as the adversary can understand RPL messages and launch attacks at any time. On the other hand, the in-threat period for PSM can be either:

- **Infinity** for all internal adversaries or external adversaries of replay/identity-cloning attacks. The former can decrypt the whole control message with the preinstalled encryption key at any time, while the latter can identify RPL control messages through the "Type" and "Code" fields of the ICMPv6 header, then replay them at any other time without the need to decrypt the message contents.
- **Zero** for external adversaries of attacks that require a full grasp of RPL control messages; e.g., rank or version attacks, due to the lack of the used encryption key.

Due to the enhanced security caused by using intra-flow NC, CSM significantly reduces the in-threat period to either:

- **The time period to receive the first UC message** for all internal adversaries (e.g., replay and identity-cloning). During this period, the adversary will wait for the first UC control message (which will be encoded with zeros and has the SC values for both UC and MC flows), so it can use the included SC values to decode any following message from any message flow. After that, it decrypts the message with the preinstalled encryption key.
- **Zero** for all external adversaries, due to the lack of the used encryption key and the correct SC values.

*3) The Necessity of Proper Recovery Mechanism:* For any message flow (UC or MC), once a message is lost, all the subsequent messages in that flow will be discarded due to the message chain breakage. This could lead to a disruption in the routing topology and sub-optimal routes. On the other hand, exchanging the missing SC values in clear text would hinder the enhanced security of CSM and allows adversaries to acquire the SC values and launch their attacks. Hence, a proper recovery mechanism that assures secure exchange for the missing SC values is needed.

## V. CONCLUSION

In this paper, we proposed a novel and new secure mode for RPL, the CSM, that is based on the concept of intra-flow NC, to enhance RPL security and to build a mitigation capability of replay attacks into the protocol itself, without significantly changing the way RPL works. A proof-of-concept prototype of CSM was devised, and its security and performance were evaluated against the currently implemented secure modes of RPL (UM and PSM, the latter with and without the replay protection mechanism) under the Neighbor attack as a demonstration. It was shown that CSM successfully mitigate replay attacks (e.g., the Neighbor attack) with less overhead and power consumption than the other secure modes. Also, it was shown that CSM has a significantly smaller in-threat period than all other secure modes. However, our evaluation indicated a need for a proper recovery mechanism for message chain breakage situations, which will be our next step.